\documentstyle[12pt]{article}

\newcommand{\be}{\begin{equation}}
\newcommand{\ee}{\end{equation}}
\newcommand{\ba}{\begin{eqnarray}}
\newcommand{\ea}{\end{eqnarray}}
\newcommand{\ft}{\footnote}
\newcommand{\nl}{\newline}
\newcommand{\al}{\alpha}
\newcommand{\bt}{\beta}
\newcommand{\ga}{\gamma}
\begin{document}

\begin{flushright}
QMW-PH-97-1
\end{flushright}

\begin{center}

\Large{\bf A Mirror Pair of Calabi-Yau Fourfolds in
Type II String Theory.}

B.S.Acharya\ft{r.acharya@qmw.ac.uk \nl Work Supported by PPARC.},

{\it Department of Physics, Queen Mary and Westfield College, Mile End Road, London. E1 4NS. U.K.}

\end{center}

\begin{abstract}
We give some simple examples of mirror Calabi-Yau fourfolds in
Type II string theory. These are realised as toroidal orbifolds.
Motivated by the Strominger, Yau, Zaslow argument
we give explicitly the mirror transformation which maps
Type IIA/IIB on such a fourfold to Type IIA/IIB on the mirror.
The mirrors are related by the inclusion/exclusion of discrete torsion.
Implicit in the result is a confirmation of mirror symmetry to genus
$g$ in the string path integral.
Finally, by considering the relationship
between $M$-theory and Type IIA theory, we show how in $M$-theory
on mirror Calabi-Yau fourfolds, mirror symmetry exchanges the Coulomb branch
with (one of the) Higgs branches of the theory. This result 
is relevant to duality in $N=2$ supersymmetric gauge theories
in three dimensions.

\end{abstract}
\newpage
\section{Introduction.}

Recent work of Strominger, Yau and Zaslow \cite{syz} has provided a deep
physical insight into the nature of mirror symmetry for Calabi-Yau $3$-folds.
This has already had an impact on the mathematical understanding of mirror
symmetry \cite{mathmir}. The SYZ argument was generalised in \cite{oz} to
the case of $n$-folds. The upshot of the argument is that
mirror symmetry is $T$-duality.

The simplest and perhaps best understood example of
a mirror pair of Calabi-Yau $3$-folds in string theory
was constructed some time ago in \cite{fqi}. These mirror target spaces
are ${T^6}/{{Z_2}\times{Z_2}}$ orbifolds and are related
by the inclusion or exclusion of discrete torsion.
This pair of string target spaces were further analysed in \cite{dis}, where
it was explicitly verified that they are indeed mirrors 
(This example was discussed again
in light of the SYZ argument in \cite{oog}). The mirror pair of target spaces
considered in \cite{dis} are very simple toroidal orbifolds. The purpose of this paper is to consider the SYZ argument applied to string theory with
Calabi-Yau fourfold target spaces which are also realised as
toroidal orbifolds. This will provide us with a strong confirmation of the SYZ
argument.

In fact, all the string target spaces that we consider in this paper occur
in the moduli space of the fourfolds constructed (in classical geometry)
in \cite{bor}. Even though the Hodge numbers of these target spaces have been
presented in \cite{bor}, we will compute them explicitly here using string
theory techniques.

The main result of the paper is that we find a strong confirmation of the
SYZ argument as well as gaining concrete examples of mirror symmetry for
Type II strings on Calabi-Yau fourfolds. The methodology follows that of
\cite{dis}, and the final result is similar in that we obtain an explicit
description of the mirror symmetry transformation of the string theory as
well as a confirmation of the symmetry to genus $g$ in the string path
integral. The key point throughout is that the ($T$-duality)mirror
transformation turns on discrete torsion in the mirror theory, which results
in the (mirror) difference between the Hodge numbers of the target spaces.
Similar examples to those studied here, but with target spaces of exceptional
holonomy were considered in \cite{ba1}.

The organisation of the paper is as follows. In the next section we consider
Type II strings on the simplest Calabi-Yau fourfold (orbifold), without
introducing any discrete torsion. We compute the Hodge numbers of this target
space. Then, using the SYZ argument, we motivate the $T$-duality
transformation that we propose implements the mirror transformation
of this theory. Using this transformation in section four we show
(following \cite{dis}) that this transformation has the effect of turning on
a certain ``amount'' of discrete torsion. We compute the Hodge numbers of the
dual target space thus verifying that the transformation is indeed the mirror
transformation. We then consider all the other possibilities of turning on
discrete torsion in this class of string target spaces. These six
possibilities result in three additional pairs of mirror fourfold
target spaces, all of whose Hodge numbers are identical and self-mirror.

Finally, we consider the relation between Type IIA theory on mirror Calabi-Yau
fourfolds and $M$-theory on the same. This leads to a mirror pair of
three-dimensional $N=2$ theories in which the Coulomb branch is
exchanged with one of two Higgs branches and vice-versa.

Mirror symmetry in higher dimensions was first studied in \cite{mird}.
Various aspects of string theory dualities in Calabi-Yau fourfold
compactifications have been discussed in \cite{go,Wit,Mayr,be,4fold}.
     
\section{Type II Strings on a Calabi-Yau Fourfold, $M$.}

Consider the 8-torus, $T^8$ with coordinates $x_{1}, x_{2}, ...x_{8}$.
The toroidal identifications on the $x_i$ are $x_i$$=$$x_{i}+1$.
Define the discrete isometry group, $\Gamma$, with generators
$\alpha,\beta,\gamma$ as follows.
\ba
&\al:x_i = -x_1,-x_2,-x_3,-x_4,x_5,x_6,x_7,x_8 \\
&\bt:x_i = x_1,x_2,x_3,x_4,-x_5,-x_6,-x_7,-x_8 \\
&\ga:x_i = -x_1,-x_2,x_3,x_4,-x_5,-x_6,x_7,x_8 
\ea

We now consider Type II string theory defined on the
above orbifold. We wish to compute the Hodge numbers
for this orbifold using the standard techniques \cite{orb}. The
Hodge numbers for this orbifold were presented in \cite{bor} and
various aspects of string/$M$/$F$-theory compactification on this
orbifold were described in \cite{go}.
However it is instructive for our purposes to compute them
explicitly here.

We will first compute the Betti numbers from which we can
compute the Hodge numbers.

To do this we need to calculate the Betti numbers in the untwisted
sector and then the twisted sector. It is straightforward
to compute that in the untwisted sector the Betti numbers\ft{These 
are simply given by the numbers of independent harmonic forms on $T^8$
that are invariant under $\Gamma$.}are
$b_{0}=b_{8}=1$, $b_{1}=b_{8}=b_{3}=b_{5}=0$,
$b_{2}=b_{6}=4$ and $b_{4}^{+}=b_{4}^{-}=11$. Here the plus/minus signs
indicate the number of harmonic four forms which are self/anti-self dual
under the eight dimensional Hodge star operator. We now turn to
the twisted sector.

In general (for abelian orbifolds), there exists a twisted sector
for each element of the orbifold isometry group which acts with
fixed points. In our example, it is clear that all seven non trivial
elements of the group act with fixed points. Of these seven, six
invert four coordinates of the torus and one ($\al\bt$) inverts all
eight coordinates. This means that the $\al\bt$ twisted sector is
$256$-fold degenerate and the other six twisted sectors are $16$-fold
degenerate. In fact, careful consideration of these latter six twisted
sectors shows that they each contribute equally to the Hodge numbers.
For this reason we will only consider one of these six and multiply
its contribution by a factor of six. 

Let us compute one of these six twisted sectors. For definiteness
we will consider $\al$. The element $\al$ acts on $T^8$ and fixes
$16$ four-tori, $T^4$. These have coordinates $x_{5},x_{6},x_{7},x_{8}$.
Each $T^4$ has its Betti
numbers given by $(b_{0},b_{1},b_{2},b_{3},b_{4})$ = $(1,4,6,4,1)$.
However, because $\al$ inverts four toroidal coordinates, the shift
in zero point energies means that in the full orbifold these $b_i$
contibute to $b_{i+2}$. This would give the full contribution
of $\al$ to the Betti numbers, {\it if} $\al$ were the only
non-trivial element in the group. However, as this is not
the case, we must project onto states which are invariant under
the whole group. 

To do this we need only consider how all the group elements act
on the coordinates of the fixed four-tori (ie $x_{5}, ...x_{8}$) and
project out the elements of cohomology which are not invariant.
It is then straightforward to compute that the $\Gamma$ invariant
harmonic forms of each fixed four-torus
contribute $(1,0,2,0,1)$ to $(b_{2},b_{3},b_{4},b_{5},b_{6})$.
As $\al$ fixes sixteen four-tori, its final contribution to the spectrum
of Betti numbers is $(16,0,32,0,16)$. As we noted above,
there are five additional twisted sectors which make an identical
contribution to the spectrum. Summing these together gives
a total contribution of $(96,0,192,0,96)$ to $(b_{2},...b_{6})$, where we
have taken into account the shift in zero point energy.

The last twisted sector is that associated with $\al\bt$. $\al\bt$
acts with $256$ points on $T^8$.
%Before GSO projection, because
The cohomology of each of these points consists of a single zero form.
However, because $\al\bt$ inverts eight coordinates, these appear as
four-forms in the cohomology of the orbifold. In fact, these states
contribute $256$ to $b_{4}^+$ (This is essentially because
the cohomology of each fixed point is identical and of a
definite parity). Adding all these contributions
we find that the spectrum of Betti numbers is

\ba
b_0 = b_8 = 1 \\
b_1 = b_7 = 0 \\
b_2 = b_6 = 100 \\
b_3 = b_5 = 0 \\
b_{4}^{+} = 363 \\
b_{4}^{-} = 107 
\ea

Let us define complex coordinates $(z_{i})$ on the orbifold as follows.
\be
z_i = x_i + ix_{i+1}
\ee
where $i$ = $1,3,5,7$.

From the definition of the orbifold, it is clear that the holomorphic
4-form is preserved. The target space of the string is therefore a
Calabi-Yau fourfold (orbifold). Moreover, using the formula for
the Dirac index A on an 8-manifold,
\be
24A = -1 + b_1 - b_2 + b_3 + b_{4}^{+} - 2b_{4}^{-}
\ee
we find that $A$=$2$; using theorem $C$ of \cite{J3}, the fourfold
has holonomy precisely $SU(4)$.

For a Calabi-Yau fourfold with $SU(4)$ holonomy, we can write the
Hodge numbers, $h^{p,q}$ in terms of the Betti numbers as follows.

\ba
&b_{0} = h^{0,0} = 1 \\
&b_2 = h^{1,1} \\
&b_3 = 2h^{2,1} \\
&b_4 = b_{4}^{+} + b_{4}^{-} = 2(h^{4,0} + h^{3,1}) + h^{2,2} = 2 + 2h^{3,1}
+ h^{2,2}
\ea

We can also use the fact that the number of real metric moduli
associated to metrics of $SU(4)$ holonomy is 
\be
b_{4}^{-} + 1 =  h^{1,1} + 2h^{3,1}
\ee

These are the numbers of Kahler and complex structure deformations which
preserve the Calabi-Yau conditions.

From these relations, and our computation of the Betti numbers it
follows that
\ba
h^{1,1} = 100 \\
h^{2,1} = 0 \\
h^{3,1} = 4 \\
h^{2,2} = 460 \\
h^{0,0} = h^{4,0} = 1 
\ea

This is in agreement with \cite{bor}. 

\section{The (T-duality) Mirror Transformation.}

In the next section, we will $T$-dualise the orbifold of the last section.
This transformation will coincide with the mirror transformation.
The $T$-duality transformation we will use will be motivated by the
Strominger, Yau, Zaslow argument \cite{syz}
concerning the implications of quantum mirror
symmetry for Calabi-Yau threefolds, generalised to the case of fourfolds
\cite{oz}. The SYZ argument implies that any Calabi-Yau $n$-fold for which
a mirror exists admits a supersymmetric $T^n$ fibration. Mirror symmetry is then just $T$-duality on the $T^n$ fibers. For us, this means 
that if we consider Type IIA or IIB string theory defined on $M$, and assume
the existence of a mirror target space ($M^\prime$),
$R\rightarrow1/R$ $T$-duality
on the $T^4$ fibers, will map us back to the IIA or IIB theory (respectively)
on $M^{\prime}$.

Since $M$ is defined as the target space of the string theory on
an orbifold, the $T$-duality/mirror transformation must be definable
on the classical orbifold ${T^8}/{\Gamma}$. This then leaves us with the
problem of identifying which $T^4$ submanifold of ${T^8}/{\Gamma}$ to
$T$-dualise. To solve this problem, we need the definition of the
supersymmetric four-torus itself.

In supersymmetric compactification
of any superstring theory or $M$-theory on some manifold $S$, the
supersymmetric cycles are identified with (homology) volume minimising
cycles. This is because if we consider branes wrapping some cycle, $D$ in
$S$, the mass of the state which appears in the non-compact
directions is proportional to the volume of $D$. If the cycle has minimal
volume, we get a state of minimum mass and hence the states corresponding
to minimal volume cycles are (supersymmetric) BPS states.
The mathematical theory of minimal
submanifolds was explored in \cite{cal}. There it was shown that if $S$
admits some globally defined form ($K$), and $K$ restricted to $D$ is 
proportional to the volume form of $D$, then $D$ is volume minimising
within its homology class\footnote{Clearly, for this to be so, the degree
of $K$ must be the dimension of $D$.}. $D$ is then said to be callibrated
by $K$. The conditions for supersymmetry
for branes wrapped around cycles have been analysed in \cite{Beck,D,oog,oz}.
There it has been shown that
(for compactification manifolds which preserve supersymmetry), the 
supersymmetric cycles are precisely the callibrated submanifolds in
\cite{cal}.

For Calabi-Yau $n$-folds with mirrors,
the supersymmetric $n$-tori which fiber the whole manifold are callibrated
by $Re(\omega)$, where $\omega$ is the holomorphic $n$-form. In the case
we are considering, $\omega$ is of degree four, so the supersymmetric
cycles callibrated by $Re(\omega)$ are four-cycles. The four-cycle we
are interested in must be a four-torus as follows from the SYZ argument
\cite{syz}. Moreover, because we have the
freedom to fix the moduli in the twisted sector, this four-torus should
exist in ${T^8}/{\Gamma}$ viewed as an orbifold in classical geometry.
The four-torus is therefore a four-cycle in $T^8$ which is preserved by
$\Gamma$. As we have noted in the previous section, there are $22$ invariant
four-cycles, and one of these is the one which interests us. Since we
will be considering $T$-duality on ${T^8}/{\Gamma}$, viewed as a complex
orbifold with complex coordinates ${z_i}$, we can appeal to what we understand
about $T$-duality and mirror symmetry for complex tori\ft{This was reviewed
in \cite{dis}.}.

For a complex torus with coordinates $z_i$, mirror symmetry just corresponds
to $R\rightarrow1/R$ $T$-duality on the circles with coordinates $Im({z_i})$.
It is natural to expect that precisely this transformation applies to
${T^8}/{\Gamma}$.
This was the case for the Calabi-Yau threefold orbifold considered
in \cite{dis}. We are therefore lead to consider the four-torus in
${T^8}/{\Gamma}$ with real coordinates $x_{2},x_{4},x_{6},x_{8}$. It is
gratifying to see that $Re(\omega)$ restricted to this cycle is the
volume form. It is clear then how to proceed. We will consider Type IIA
(or IIB) string theory defined on the orbifold $M$ of the previous section
and consider $R\rightarrow1/R$ $T$-duality transformations in the $2,4,6,8$ 
directions. This should map us to Type IIA (or IIB) theory on a target
space with Hodge numbers mirror to those of $M$. We will now
verify that this is the case.

\section{Type II Strings on the Mirror of M.}

We have been lead to consider $T$-dualising the circles in ${T^8}/{\Gamma}$
with coordinates $({x_{2},x_{4},x_{6},x_{8}})$. Let us refer to this
transformation as $T$. As was pointed out in \cite{dis}, $T$-duality
for orbifolds such as ours can result in the turning on of discrete
torsion \cite{vaf} in the $T$-dual orbifold. In \cite{dis}, it was discrete
torsion that was responsible for ``producing'' the mirror of a Calabi-Yau 
threefold. This will be the case here. However, in the case considered
in \cite{dis}, the orbifold isometry group was ${Z_{2}}\times{Z_{2}}$ and hence
the discrete torsion was $Z_2$ valued\ft{The group in which the discrete
torsion takes values corresponds to $H^{2}(G,U(1))$ \cite{vaf}, where
$G$ is the orbifold group.}. Thus, in \cite{dis}, the discrete torsion
was unique. For our case, the discrete torsion is not unique
and hence there exist several possibilities. In fact there exist seven
non-trivial modular invariant possibilities for what the discrete torsion may
be. However, we will see that
under the transformation $T$, just the right ``amount'' of discrete torsion
will be present in the $T$-dual orbifold to produce the mirror of $M$. We will
now derive the discrete torsion phase factor which appears in the path integral
measure when we apply the transformation $T$ to the string theory on $M$. The
discussion follows \cite{dis} which we will briefly review here.

Consider the $n$-torus with coordinates $({x_1},{x_2},...{x_n})$. Superstrings
propagating on the torus have coordinates $X_i$ with fermionic partners ${\psi}^i$. Under an $R\rightarrow1/R$ $T$-duality transformation in any one of the
$n$-directions, the measure of the genus $g$ path integral for the theory,
${\mu}_g$ transforms as follows.
\be
\mu_g \longrightarrow {(-1)^{{\sigma}_{\alpha}}}\mu_g
\ee
where
\be
\alpha = ({{\theta}_i},{\phi}_j)
\ee
is the spin structure of the genus $g$ Riemman surface (relative to the
canonical $({a_i},{b_j})$ basis of $1$-cycles) and
\be
\sigma = \theta.\phi
\ee
is its parity. We can write ${\theta}_i$=$\Theta$, a $g$-vector and similarly
for $\phi$. If we $T$-dualise $k$ directions, the factor of ${(-1)^{\sigma}}$
appears $k$ times. These factors are responsible for the $T$-duality of
Type IIA/IIB string theories compactified on $n$-tori.

In the orbifold that we are discussing the above transformation rule
gets modified due to the twist by $\Gamma$ in the following way.

With the generators of $\Gamma$ as we have defined them, a general twist
of the genus $g$ Riemman surface in the $({a_i},{b_j})$
directions involves a pair of elements that we
can write as $({{\al}^R}{{\bt}^S}{{\ga}^T},{{\al}^U}{{\bt}^V}{{\ga}^W})$.
Here, $R,S,T,U,V,W$ are $g$-vectors which describe the change in boundary
conditions of the ${\psi}_i$ when going around the $1$-cycles on the genus
$g$ world sheet. Under a $T$-duality transformation in the $r$th direction
the measure will pick up a factor corresponding to a shifted spin structure.
For example, if we begin with Type IIA theory on our original orbifold and
$T$-dualise the circle with coordinate $x_8$, the spin structure gets
shifted as follows:
\be
({\Theta},\Phi) \longrightarrow (\Theta + S,\Phi + V)
\ee
This is because $\bt$ is the only generator which acts on $x_8$ and its action
is just a minus sign.
The genus $g$ measure therefore picks up a factor which is $(-1)$ raised
to the parity of
this shifted spin structure.

In the problem we are addressing, we are dualising the $(2,4,6,8)$ directions
and all we need to do now is calculate all the factors which appear in the
$T$ transformed path integral. These are the parities of the shifted spin
structures in the dualised directions. These are:
\ba
&{x_2}: \sigma(\Theta + R + T,\Phi + U + W)\\
&{x_4}: \sigma(\Theta + R,\Phi + U)\\
&{x_6}: \sigma(\Theta + S + T,\Phi + V + W)\\
&{x_8}: \sigma(\Theta + S,\Phi + V)
\ea
Each of these factors appears as a power of $(-1)$ in the $T$ transformed
path integral measure. A useful formula for shifted spin structures is
\cite{at}:
\be
\sigma(\Theta+A+B,\Phi+C+D)=\sigma(\Theta,\Phi) + \sigma(\Theta+A,\Phi+C)
+ \sigma(\Theta+B,\Phi+D) + A.D - B.C
\ee

Using this we find that the path integral measure transforms under $T$ as
\be
\mu_g \longrightarrow (-1)^{R.W - T.U + S.W - T.V}.\mu_g
\ee
This formula shows that under $T$, the IIA/IIB theory on the orbifold with
generators $\Gamma$ transforms into the
IIA/IIB theory with generators $\Gamma$ but with discrete torsion (given
by the above factor) turned on. We will call the $T$-dual target space
$M^{\prime}$.

\subsection{The Hodge Numbers of $M^{\prime}$.}

In this subsection we will calculate the Hodge numbers for the
string theory on $M^{\prime}$. Discrete torsion has no effect on
the untwisted sector of an orbifold. The contribution from this
sector is therefore the same as in section 2.

Let us consider the $\alpha$ twisted sector. As before, $\al$ fixes
sixteen 4-tori. We then have to project on to states invariant under
$\Gamma$. This is where the effect of discrete torsion comes into being.
In general, certain states which would have survived the $\Gamma$ projection
in the absence of discrete torsion will now remain invariant (and vice-versa).
This is most easily seen in the Hamiltonian framework \cite{dis}. Let us
consider the $\sigma$ and $\tau$ directions of the world sheet as space
and time. As we have noted, the general twist of the world sheet
in these directions is given by $({{\al}^R}{{\bt}^S}{{\ga}^T},
{{\al}^U}{{\bt}^V}{{\ga}^W})$. For a fixed $\tau$, we
have a Hilbert space of states consisting of all the possible twisted
states associated with $\Gamma$. However, as $\tau$ evolves, because
of the projection by $\Gamma$ in these directions also, we project
onto $\Gamma$ invariant states. For example, the element $({\al,\bt})$ 
corresponds to a sector of the Hilbert space consisting of the $\al$
twisted sector projected onto $\bt$ invariant states. If, for this
element the discrete torsion factor is $(-1)$, then we include an
extra minus sign in the action of $\bt$ on the $\al$ twisted sector, relative
to its ``geometric'' action in the absence of the torsion.

Let us consider the $\al$ twisted sector. This corresponds to
elements of $\Gamma$ of the form $(\al,{{\al}^U}{{\bt}^V}{{\ga}^W})$.
Our discrete torsion in a general sector is of the form
\be
\epsilon = (-1)^{R.W - T.U + S.W - T.V}
\ee
Thus, in the $\al$ twisted sector the discrete torsion factor is
\be
\epsilon = (-1)^{W}
\ee

From this it follows that we have to insert an extra minus sign in the
action of elements of $\Gamma$ which contain an odd number of $\gamma$'s when
acting on the cohomology of the sixteen 4-tori fixed by $\al$. These
elements are $(\ga,\al\ga,\bt\ga,\al\bt\ga)$. The remaining four elements
act with their standard geometric action as if the discrete torsion
was not present. Putting these facts together, it is straightforward
to compute that in the orbifold with discrete torsion, the contribution
of each 4-torus fixed by $\alpha$ to the Betti numbers is to add
$(0,0,4,0,0)$ to $({b_2},{b_3},{b_4},{b_5},{b_6})$\ft{If we denote
by $x_{ij}$ the two-form, ${dx_i}\wedge{dx_j}$ on the fixed $T^4$, then
the two-forms $x_{ij}$ for $(ij)=[(57),(58),(67),(68)]$ are invariant.
Because of the zero point energy in the twisted sector, these contribute
to the spectrum of harmonic four-forms on $M^{\prime}$.}. Moreover, of these
four-forms, two are self-dual/anti self-dual. This gives a total
contribution to the Betti numbers in the $\al$ twisted sector of
$32$ self-dual and $32$ anti self-dual four forms.

Considering carefully the other five twisted sectors which are sixteen-fold
degenerate, we find that these contribute an identical spectrum to that
of $\al$. However, we would like to stress that this is a highly non-trivial
fact and is not dictated by any obvious symmetry arguments and needs to
be checked explicitly.

Finally, we need to consider the $\al\bt$ twisted sector. This corresponds
to elements of the form $(\al\bt,{{\al}^U}{{\bt}^V}{{\ga}^W})$, for
which it is straightforward to see that the discrete torsion is trivial.
This sector therefore makes the same contribution to the spectrum as
we found in section 2, which is $256$ to $b_{4}^{+}$.

Combining all these contributions we find that the relevant Betti numbers of
$M^{\prime}$ are:

\be
(b_2,b_3,b_{4}^{+},b_{4}^{-}) = (4,0,459,203)
\ee
Using the formulas of section 2, we find that $A=2$ and
\be
(h^{1,1},h^{2,1},h^{3,1},h^{2,2}) = (4,0,100,460)
\ee

We can therefore see that $h^{p,q}(M) = h^{4-p,q}(M^{\prime})$. Hence
the Hodge numbers of the two target spaces are mirror to one another.
We conclude that $T$ is indeed the mirror symmetry transformation.

\subsection{Further Examples.}

As we pointed out in the previous subsection, there exist eight a priori
different orbifolds with generators $(\al,\bt,\ga)$ given in equations
$(1)-(3)$. This is due the fact that in addition to the orbifold without
discrete torsion, we can turn on seven non-trivial discrete torsion factors
consistent with modular invariance. In fact the discrete torsion group in this
case is ${Z_2}\times{Z_2}\times{Z_2}$, generated by $[{(-1)}^{R.V-U.S},
{(-1)}^{R.W-T.U},{(-1)}^{S.W-T.V}]$.
It is natural to consider the mirror symmetry
transformation $T$ in all these cases. Let us define the three generators
of the discrete torsion group as $({g_1},{g_2},{g_3})$ respectively. Let us
also denote the target space of the orbifold theory with discrete torsion
factor $k\equiv$
$({{g_1}^a}.{{g_2}^b}.{{g_3}^c})$ by $M_k$. In this notation, the orbifold
without discrete torsion and its mirror are denoted by $M_1$ and
$M_{{g_2}.{g_3}}$ respectively. 

It is clear from the discussion of the previous subsection that under
$T$, any of the eight target spaces $M_k$ will transform under $T$ to
another of the $M_k$. In fact $T$ produces four pairs of target spaces, which
in principle are mirrors of one another, ie the eight $M_k$ ``transform''
as four doublets under $T$. It is also clear that, under $T$,
the target spaces $M_k$ transform as follows:

\be
{M_k} \longrightarrow M_{k.{g_2}.{g_3}}
\ee

The four pairs of target spaces related by $T$ are therefore given
by pairs $(k,k.{g_2}.{g_3})$. With the expectation that these pairs of
target spaces 
should be mirror to one another we can compute the Hodge numbers for each pair
to verify this. This gives the result that apart from the mirror pair (with
$k=1$) discussed above, the Hodge numbers for all other pairs ($k\neq1$)
are identical and self-mirror. In fact the Hodge numbers of $M_k$, for
$k\neq1,{g_2}.{g_3}$ are:

\be
({h^{1,1}},{h^{2,1}},{h^{3,1}},{h^{2,2}}) = (20,64,20,76)
\ee

Since these are self-mirror, we get no contradiction with the SYZ argument
that lead us to conclude that $T$ is the mirror transformation. We would also
like to note that for reasons explained in \cite{dis}, implicit in the above
result is the fact that mirror symmetry between the dual pairs of
theories is confirmed to genus $g$ in the string path integral.

\section{Relation to $M$-theory.}

In this section we briefly discuss the implications of mirror symmetry
for Calabi-Yau fourfolds for $N=2$ vacua of $M$-theory in three dimensions.
$M$-theory compactified on a manifold of $SU(4)$ holonomy gives a three dimensional theory with $N=2$ supersymmetry. General aspects of such
compactifications were studied in \cite{be}, whereas the relationship
of such vacua (via $F$-theory) to $N=1$ vacua in four dimensions has been
studied in \cite{Wit,Mayr}. 

Type IIA theory compactified on a Calabi-Yau
fourfold $(M)$ has a strong coupling limit (keeping the Calabi-Yau moduli fixed)
which is the three dimensional theory obtained from $M$-theory on $M$.
Moreover if we now consider IIA theory on $M^{\prime}$, the mirror of
$M$, then it has a limit\footnote{In \cite{phase} it was argued that $M$-theory
is devoid of exotic phase transitions to compactifications which are defined
in the corresponding IIA theory as abstract conformal field theories
with no definite analogue in classical geometry. We will assume
in this section that the mirror target spaces are well defined in
classical geometry.} which is $M$-theory on $M^{\prime}$. This chain of
dualities provides a duality between $M$-theory on $M$ and $M^{\prime}$.
Let us discuss what this implies for the low energy spectrum.

In uncompactified $M$-theory the low energy dynamics is 
described by eleven dimensional supergravity. The relevant
degrees of freedom are the 3-form potential $A_3$ and the metric $G$.
Let us consider the low energy spectrum of $M$-theory compactified on
$M$ and its mirror. We will denote by $h_{p,q}$ the Hodge numbers of $M$
and by $h^{\prime}_{p,q}$ those of $M^{\prime}$. As the two manifolds are
mirror, $h_{p,q}$ = $h^{\prime}_{4-p,q}$.

In $M$-theory on $M$, the three form contributes $h_{1,1}$ massless vector fields
and $2h_{2,1}$ scalars. The metric contributes $h_{1,1}$ scalars (Kahler
moduli) and $2h_{3,1}$ further scalars (complex structure moduli). All these
massless fields combine into the following multiplets of $N=2$ supersymmetry.
The $h_{1,1}$ vectors and scalars combine to give $h_{1,1}$ vector multiplets, 
$V_i$. The scalars from $A_3$ form $h_{2,1}$ scalar multiplets, $S_s$. The 
complex structure moduli from the metric form $h_{3,1}$ additional scalar multiplets, $H_h$. Vectors in three dimensions are dual to scalars so we
can dualise the vectors to form $h_{1,1}$ scalar multiplets which we also
denote by $V_i$. The $V_i$ parametrise a $2h_{1,1}$ dimensional Coulomb
branch, $V$. The other scalars parametrise two Higgs branches, $H_1$ and
$H_2$ of respective dimensions $2h_{3,1}$ and $2h_{2,1}$.

In $M$-theory on $M^{\prime}$ we have a $2h_{3,1}$ dimensional Coulomb branch,
$V^{\prime}$, a $2h_{1,1}$ dimensional Higgs branch ${H_1}^{\prime}$ and
an additional $2h_{2,1}$ dimensional Higgs branch, ${H_2}^{\prime}$.
Mirror symmetry of the Type IIA theory on $M$ and $M^{\prime}$ implies
a mirror symmetry between $M$-theory on $M$ and $M$-theory on $M^{\prime}$.
By a simple counting of the moduli space dimensions of the various
branches it is straightforward to see that mirror symmetry is the following
exchange symmetry:

\ba
&V \leftrightarrow {H_1}^{\prime}\\
&{H_1} \leftrightarrow {V}^{\prime}\\
&{H_2} \leftrightarrow {H_2}^{\prime}\\
\ea 

This result has obvious implications for duality in $N=2$ gauge theories
in three dimensions. Examples of such gauge theories have recently been
discussed in \cite{boer}. It would be interesting to explore this further.

\section{Discussion.}

In the examples of mirror symmetry that we have discussed in this paper,
it is clear that the Strominger-Yau-Zaslow
argument \cite{syz} (or its generalisation to $n$-folds \cite{oz}) is a
very powerful one. In turn we hope that the mirror pairs of string theories
constructed here, being so simple, can be used to check other properties
of mirror symmetry in higher dimensions \cite{mird}, and may even provide
insights into other properties of higher dimensional mirror symmetry.

The relationship between mirror symmetry for fourfolds and $N=2$
$M$-theory compactifications is also very promising for future investigations.

\section{Acknowledgements.}

The author would like to thank M.Atiyah for discussions concerning spin
structures and J.M.Figueroa-O'Farrill for discussions concerning discrete
torsion. He would also like to thank the PPARC, by whom this work is
supported.

\end{document}